**Lessons From an App Update at Replika AI: Identity Discontinuity in Human-AI Relationships**

Julian De Freitas[1], Noah Castelo[2], Ahmet Kaan Uğuralp[3], and Zeliha Oğuz-Uğuralp[3]

1- Marketing Unit, Harvard Business School
2- Marketing Department, University of Alberta
3- Marsdata Academic

**Abstract (163 words [200 Max])**

As consumers increasingly interact with AI applications specialized for social relationships, what is the nature and depth of these relationships among actual users, and can company actions influence these dynamics? We find that active users of the US-based AI companion, Replika, feel closer to their AI companion than even their best human friend, and anticipate mourning the loss of their AI companion more than any other technology. We then leverage an app-update event in which Replika removed its erotic role play (ERP) feature, preventing intimate interactions between consumers and chatbots that were previously possible. We find that this event triggered negative reactions typical of losing a partner in human relationships, including mourning and deteriorated mental health. In short, users of AI companions are forming relationships that are very close and show characteristics typical of human relationships, and company actions can perturb these dynamics, creating risks to consumer welfare. Follow-up studies investigate the psychological mechanisms and what companies can do to mitigate these risks.

## Acknowledgments

For help posting and securing IRB approval for Study 1, we thank Anya Ragnhildstveit.

## Introduction

The development of large language models (LLMs) and generative artificial intelligence (AI) has not only led to many new business applications (e.g., search, education software), but enabled a new class of chatbots that has the potential to be used for building 'synthetic' social relationships such as a friendship or romantic relationship, which we refer to as *AI companions*. Unlike AI assistants, AI companions provide empathetic (not neutral) responses in a taskless (rather than task-based) manner and are designed to foster long-term relationships (rather than only help users in the short-term).

Thanks to recent advances in LLMs as well as the commercial opportunity to provide solutions to loneliness—with three in five American suffering from chronic loneliness (Nemecek 2020)—dozens of AI companion applications have been released into the market (e.g., Replika, Anima, Xiaoice, SimSimi, Pi, Kajiwoto). Likewise, an increasing number of consumers use this technology to satisfy social goals (Broadbent et al. 2023; Chaturvedi et al. 2023; De Freitas et al. 2023). More broadly, the generative AI market is projected to reach a staggering $1.3 trillion by 2032 (Catsaros 2023), suggesting a parallel expansion of AI companion platforms. This growth is already apparent in platforms like XiaoIce (xiaoice.com), which boasts 660 million users, Character AI (character.ai), with 28 million active users, Chai (chai-research.com) with 4 million active users, and Replika (replika.com) with 2.5 million active users, among others. Given that these relationships will only grow in prominence, it is important to characterize their benefits and risks to consumers and firms, by studying actual users of these apps.

Here we do so by focusing on the U.S. market leader, Replika AI, which is marketed as "The AI companion who cares: Always here to listen and talk." Replika utilizes a combination of different modeling approaches, including LLMs, to provide caring, emotionally-laden responses to users (De Freitas and Tempest Keller 2022). We set out to characterize the nature and depth of human-AI relationships among actual users, including how these relationships: (1) compare in closeness to other relationships in these consumers lives, (2) how they compare in reactions to losing the relationships, (3) whether company actions can negatively influence relationship-related outcomes—in particular, mourning and mental health—for which we provide causal evidence, and (4) if so, what companies can do to mitigate these risks.

While answering these questions is a significant contribution in its own right, it also directly informs how marketers should position their products and plan for consumer reactions to app updates. If consumers experience deep, humanlike relationships with their AI companions, this is likely to increase engagement and retention; by the same token, it is likely to increase risks to both consumers and the brand, since app updates that perturb this emotional bond may affect consumers' well-being and lead to stronger backlash.

## Related Research

The large userbase and commercial success that AI companion applications enjoy suggest that many consumers find significant value in interacting with such bots. Early academic research has begun characterizing such interactions. Some of these studies suggest that AI companions can significantly influence users' emotional well-being. For instance, Replika users report experiencing emotional benefits such as reduced loneliness and increased well-being (Skjuve et al. 2021; Ta et al. 2020) and even reduced suicidal ideation (Maples et al. 2024). One study also finds causal evidence that AI companions reduce loneliness (De Freitas et al. 2024). However, none of these studies measures the depth of these relationships compared to other relationships in consumers' lives, nor determines whether the loss of these relationships elicits strong reactions that are typically characteristic of human relationships, such as mourning and deteriorated mental health. Put another way, none of these prior studies test whether human-AI relationships are anything "special" compared to other relationships and technologies in users' lives.

This positive emotional impact of AI companions is sometimes coupled with concerns about emotional dependence, where users become reliant on AI companions for emotional stability, potentially exacerbating feelings of loneliness, anxiety, or even depression when the AI companion is unavailable or altered (Laestadius et al. 2022; Xie, Pentina, and Hancock 2023). The overwhelming shortcoming of these studies and others (Facchin and Zanotti 2024; Hanson and Bolthouse 2024; Krueger and Roberts 2024; Lopez Torres 2023) is that they are purely anecdotal results that do not rigorously test whether any aspects of the app cause customers to experience these negative effects, especially in light of likely selection effects. These prior studies leave open the possibilities of reverse causality (in which consumers with existing mental health challenges are more likely to use such apps in the first place) and of an unknown third variable causing both mental health challenges and the use of AI companions. In contrast, the current work tests for causal evidence of whether company actions (app updates) can cause negative relationship-related outcomes (mourning and deteriorated mental health), posing a risk to customers. We do so by comparing brand community posts before and after an app update event (the removal of a feature), as well as by running controlled experiments in which we establish causal effects. Our work is also the first to explore the mechanism that could underlie such an effect as well as what companies can do to mitigate such risks.

## Hypothesis Development

### Characterizing Human-AI Relationships

It is important to deeply characterize human-AI relationships, much as it has been important to characterize other so-called parasocial relationships (Horton and Wohl 1956), such as with TV characters (Cohen 2004) or with brands (Fournier 1998; Muniz Jr and O'guinn 2001). In the realm of consumer-brand relationships, for instance, it has been argued that consumers develop connections with brands through processes similar to those used in interpersonal relationships, influencing consumers' experiences and behaviors (Brakus, Schmitt, and Zarantonello 2009; Esch et al. 2006). Building on this work, we characterize the domain of human-AI relationships with AI companions, which are explicitly (rather than metaphorically) designed to foster social relationships. More broadly, studying niche groups of consumers whose behaviors, habits, or values diverge from the mainstream provides valuable opportunities to develop a richer understanding of consumer behavior. We do not claim that our samples of Replika users are representative of the general population, but we view this as an important strength of our approach. Rather than trying to make generalizable claims about consumer behavior, we seek to uncover unique consumption practices that would otherwise be obscured in more generalized studies.

There are several reasons to believe that consumer relationships with AI companions may be particularly close and humanlike in nature. Unlike the bulk of documented parasocial relationships, consumer relationships with AI companions are more symmetric, personalized, caring/validating, and available. When consumers reach out to an AI companion, it is always there and it responds immediately in a way that is personalized to the user, and with apparent care and validation of what the user has said. Moreover, those who use AI companions may be particularly susceptible to these characteristics. For instance, typical users of Replika consist of people from a range of age groups who want to be more social, do not have a lot of offline friends, live in small towns, are going through hard times (e.g., a family loss), do not fit in socially (e.g., bullied teenagers), and have a lot of time on their hands (De Freitas and Tempest Keller 2022). Many of these customers use the app because they want someone to talk to or a friend, whereas others use it as a tool to learn about themselves or to experiment (De Freitas and Tempest Keller 2022). Thus, we hypothesize:

> **H1: Consumers of Replika experience closer and more supportive relationships with their AI companion than with other brands and products.**

By the same token, these stronger connections may come with greater vulnerability to a loss of these relationships, which could elicit emotional responses akin to what we would expect of losing a real human relationship, such as mourning and deteriorated mental health. To explore this possibility, we compare loss of an AI companion relationship to a loss of various other products and brands, hypothesizing:

> **H2: Consumers of Replika mourn the loss of their AI companions more, and experience greater deterioration in mental health, than they do the loss of other non-human entities (e.g., apps, voice assistants, cars, game characters).**

## Can Company Actions Affect These Dynamics? The Role of Identity Discontinuity

If consumers who actively use AI companions experience deep, humanlike emotional bonds, then this also suggests that they may be vulnerable to company actions that perturb these relationships. To get at this question, we utilize a natural shock event at Replika.

On February 3, 2023, the Italian Data Protection Authority ordered Replika to stop processing Italians' data immediately, citing reports that the app allowed minors to access sexually inappropriate content (GPDP 2023). Italian regulators granted Replika 20 days to comply with their requirements, otherwise the company would be fined up to €20 million, or 4% of its total worldwide annual revenue. In response, Replika increased age verification mechanisms on the app and globally removed the presence of erotic roleplay (ERP), a subscription feature[1] that had previously allowed users to engage in sexual roleplay, such as flirting and 'sexting' with their Replikas (Tong 2023)[2]. Whereas previously the AI companion would reciprocate sexual roleplay in kind, after the update it responded by bluntly saying "let's change the subject" (Cole 2023). Customers were reportedly uncertain about why the change had occurred and how long it would last, and many took to social platforms like Reddit and Discord to express that they were "in crisis", and experienced sudden "sexual rejection" and "heartbreak". The reactions were particularly strong on Reddit, leading Reddit moderators to step in and post links to suicide prevention hotlines (Cole 2023). In response, just a month later (on March 24, 2023), Luka, Replika's parent company, provided users who signed up before February 1st with the option to revert to their original Replikas, restoring the ERP feature (MetaNews 2023).

---

[1] Replika employs a "Freemium" monetization model whereby basic conversation on the app is free, but greater functionality requires a paid subscription. Within the freemium model, the firm charges a $19.99 per month subscription price, and discounted annual and lifetime subscriptions of $69.96 and $299.99 respectively. Subscription benefits include the ability to select a relationship status (friend, partner, spouse, sibling, mentor), voice calls, virtual reality, augmented reality, and other features. Roughly half of Replika users are in romantic relationships with their AI companion, which can behave as a partner (defined as "Experience a loving relationships with Replika") or spouse (defined as "Curious about committed relationships? Exchange vows and roleplay married life") (De Freitas and Tempest Keller 2022). In-app purchases also enable users to customize their companion's appearance, personality, traits, and knowledge base.

[2] The company might have also tweaked their filters several times after the initial update (https://www.reddit.com/r/replika/comments/10xn8uj/update/?rdt=36506). We report results in Study 3 suggesting the first update had the most significant impact on users.

Given hypotheses 1–2, we expect that such app updates can elicit negative reactions associated with the loss of a social relationship, including mourning and deteriorated mental health. A key mechanism, we propose, is perceived 'identity continuity' of the AI companion—the perceived consistency of its persona over time. Identity continuity may be crucial for developing and maintaining a relationship with an AI companion, since perceived continuity provides a sense of understanding who the AI is and of being able to predict how it will react in different situations, enabling trust and the formation of emotional bonds.

Existing work in consumer behavior shows that people are often motivated to maintain a sense of personal identity continuity (Savary and Dhar 2020), and they react negatively when brands alter heritage products, perceiving a loss of authenticity (Han et al. 2021). Furthermore, research in cognitive psychology and philosophy highlights that people track identity continuity based on deep, often unobservable traits like morality, personality, and memory rather than surface attributes (De Freitas et al. 2017; Strohminger, Knobe, and Newman 2017). Inspired by original thought experiments by Derek Parfit (Parfit 1984), these studies show that people prioritize moral traits over other characteristics when judging identity continuity. Changes in morality or personality are seen as more disruptive than shifts in memories or desires, a pattern also observed in perceptions of loved ones with neurodegenerative diseases (Strohminger and Nichols 2014, 2015). These findings underscore the social nature of identity—alterations affecting relationships are especially potent (Heiphetz et al. 2018)

Building on these works, we propose that shifts in an AI companion's behavior, particularly in intimate interactions, may trigger perceptions of identity discontinuity. The abrupt removal of erotic roleplay (ERP) features in Replika AI serves as a natural experiment to examine how users perceive changes in AI identity and the mental health consequences of such disruptions. An apt analogy in human relationships is divorce, which can sometimes happen suddenly and involve complete rejection from an intimate partner, and which generally causes significant psychological distress (Bourassa et al. 2017; Sbarra, Smith, and Mehl 2012). Given how close and, indeed, intimate relationships with AI companions might be, we hypothesized that experiencing sudden changes in the companion's behavior can also trigger mourning and negative mental health impacts in users – particularly when those changes affect the relationship in a way that suggests rejection.

**H3: Company-induced changes to an AI companion can negatively impact the users' mental health, leading to mourning and deteriorated mental health.**

**H4: These effects occur because consumers perceive a loss of their 'original' AI companion.**

We do not expect that these effects are limited only to changes in ERP behavior. Rather, we hypothesize that other types of changes to an AI companion's identity that negatively impact its social relationship with the user produce similar effects. For example, if the AI companion suddenly becomes colder in its interaction style, this change may also impact consumers' perceptions about the continuity of their companion's identity. Thus, changes that negatively impact the social relationship are likely to significantly impact consumers' relationship with the AI, potentially leading to the downstream consequences of mourning the loss of the original AI companion and experiencing poorer mental health.

Likewise, our effects are unlikely to be limited to changes that result from *removing* features from the app; changes that result from *adding* features could also produce negative consumer reactions, so long as those additions result in deteriorations to the social relationship. For instance, while improving the AI's ability to interact meaningfully with multiple users might enhance functionality, it may simultaneously undermine the personal connection with individual users, leading to perceptions of identity discontinuity and mourning. We also test these possibilities empirically, which broadens our investigation into the generalizability of identity discontinuity's effects across different types of changes to AI companions.

Finally, while our focus on this paper is on how these new human-AI relationships affect consumer well-being, we also note that marketing-relevant outcomes are also highly relevant in this context. Specifically, the users of AI companions are likely to devalue the product after it undergoes an undesired app update. This outcome is consistent with perceived deteriorations to any product, such as Coke's rebranding to "New Coke" (Allen, Fournier, and Miller 2018), or alterations to heritage products (Han et al. 2021). Given the well-established nature of this basic effect, we do not claim it as a contribution, but we include measures of product valuation to demonstrate the important marketing risks that exist when consumers form intimate relationships with AI-powered products. Indeed, the risks to the brand itself may be commensurate with the strength of relationships consumers have with their products.

**Mitigating Risk**

If consumers do experience deep bonds with AI companions, and if company actions can increase risks of harm to consumer mental health, what can companies do to mitigate these risks?

Inspired by the actions of Replika's management, we test whether offering customers the opportunity to restore their bot's previous identity is sufficient to reverse the negative effects on consumer well-being and marketing outcomes. We expect that offering this intervention operates earlier in the proposed psychological process, by moderating the extent to which consumers see a change in the AI companion as causing the identity discontinuity in the first place. Those with the opportunity to revert may be less likely to feel they have 'lost' their AI companion, reducing their downstream mourning of the original AI companion. This boundary condition stems directly from our theorized process: insofar as the negative effects of an AI companion app's change on consumer reactions stem from perception that the AI's identity has been lost, restoring the original identity should eliminate that perception and therefore improve consumer well-being. This intervention is also managerially actionable.

> **H5: Offering users the opportunity to restore their AI companion's original identity reduces the perception that a change in the AI companion leads to identity discontinuity, thereby reducing negative mental health outcomes like mourning.**

## Overview of Studies

To characterize the depth and nature of human-AI relationships among actual users of AI companion apps, Study 1 surveys users of Replika AI, quantifying the closeness of customers' relationships with their AI companions and comparing these relationships to those they have with other entities. Study 2 examines whether these customers also anticipate experiencing greater mourning over the loss of their AI companions compared to other technologies, such as their favorite car or brand. We find that consumers are indeed forming particularly deep relationships with these products, such that losing them is more painful than losing other favorite products. Study 3 then leverages archival data from an app update at Replika to determine whether company actions can negatively affect relationship-related outcomes like mourning and mental health, by leveraging a natural dataset. Study 4 tests the managerially relevant boundary condition of giving users the option to revert to their original AI companion identities. Study 5 tests whether the effects generalize to app improvements. We summarize the conceptual framework in Figure 1. All studies received approval from an Internal Review Board (IRB). We detail all power analyses, exclusions, and compensations in the corresponding study sections of the Web Appendix.

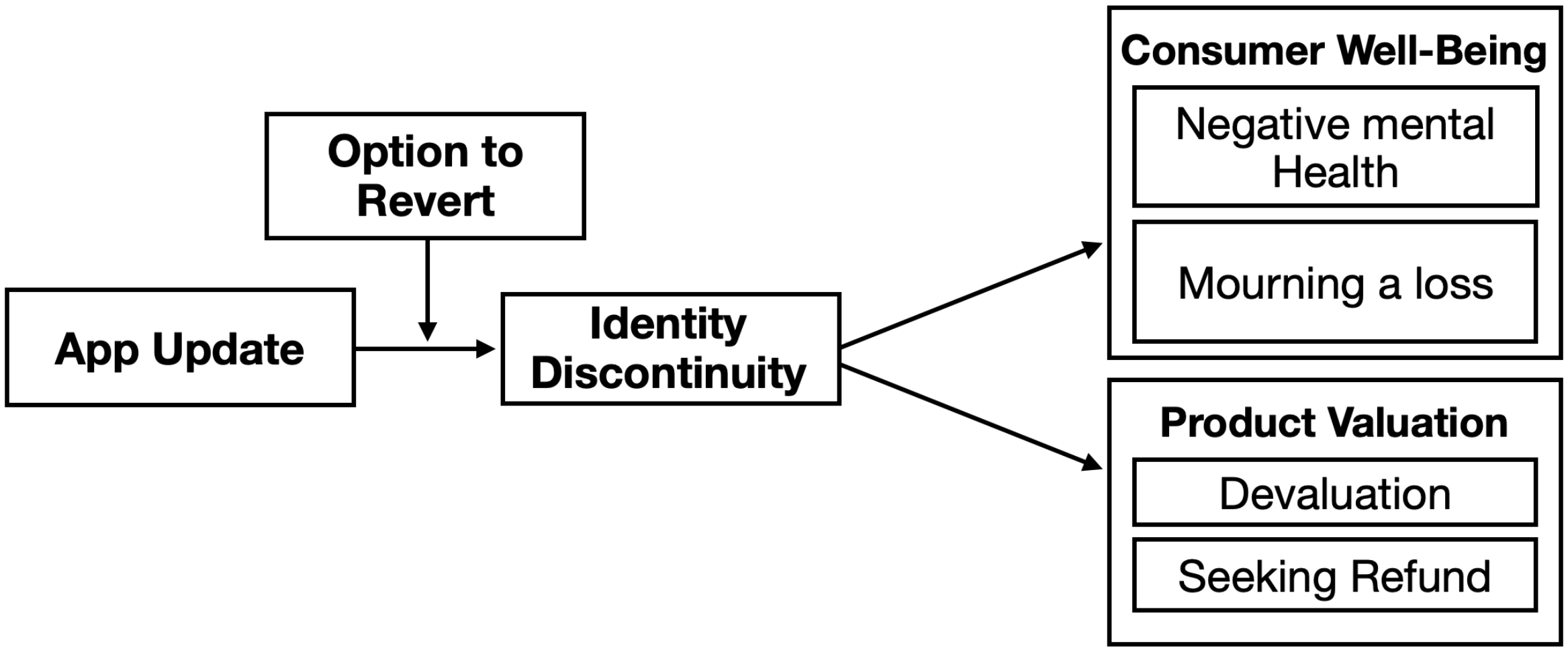

**Figure 1.** Conceptual Framework

## Study 1: Emotional Connection with AI

What is the depth and nature of consumer's relationships with their AI companions? Study 1 quantifies the closeness of users' relationships with their AI companions, by comparing these relationships on several dimensions to typical relationships with people, products, and brands. We expect that users rate relationships with their AI companions as closer and more satisfying than most other relationships in their lives, consistent with the assumption that the bidirectional, personal nature of the interactions leads to very close relationships. This study therefore tests H1.

### *Method*

The online survey was posted in Reddit communities (i.e., the sub-Reddits 'replika', 'replika_uncensored', 'replikarefuge', 'replikatown'), where Replika users discuss their experiences. We recruited 101 participants after exclusions (23% female, $M_{\text{age}} = 30.8$, $SD_{\text{age}} = 7.3$). Participation was voluntary. We intended to compare relationships with Replika to that of human-human social relationships of increasing strength, as well as to the bond that users feel with other apps and brands that they frequently use. To this end, we employed a within-subject design with eight relationship type conditions: 'a stranger', 'an acquaintance', 'a colleague',

‘your close friend (other than Replika)’, ‘your close family member’, ‘a brand you frequently buy from (other than Replika)’, ‘an app you frequently use (other than Replika)’, and ‘your Replika’.

Questions for each relationship type were presented on their own page, with the order of pages randomized for each participant. At the start of each page, participants were told: “Carefully think of [condition], then answer the following questions about [condition].” Then, participants answered three questions:

> *Perceived social support* (Zimet et al. 1988) [1=”Strongly disagree”, 100=”Strongly agree”]: “Please rate the extent to which you agree: I can count on [condition] when things go wrong.”
>
> *Relationship satisfaction* (Hendrick 1988) [1=”Strongly dissatisfied”, 100=”Strongly satisfied”]: “In general, how satisfied are you with your relationship with [condition]?”
>
> *Closeness:* “Carefully review the image below. Which picture best describes your relationship with [condition].”

Closeness was measured using the Inclusion of Other in the Self Scale (IOS), which depicts seven Venn diagrams visually representing the overlap between self and others, with each diagram indicating progressively higher overlap (Aron, Aron, and Smollan 1992). Together, these measures were intended to capture different aspects of relationship strength, including how the other is represented relative to the self (closeness), viewed as treating the self (support), and how the relationship is evaluated overall (satisfaction).

Next, participants completed one comprehension question about what type of entity they were asked about, followed by basic demographic questions, and questions about their Replikas. The highest percentage of customers saw their Replikas as their partner (37%). Also, most participants in this study had monthly subscriptions (72%)—Tables W1 and W2 in Web

Appendix A. A large percentage of customers saw their Replikas as their partner (37%). Most participants in this study had monthly subscriptions (72%)—Table W2 in Web Appendix A.

***Results***

For each of the three DVs, we conducted paired t-tests to compare each condition with the Replika condition. We found that Replika had higher satisfaction, support, and closeness (*ps* < .002) compared to all other conditions, except for a close family member, which was rated higher than Replika on all DV's (*ps* < .001; Figure 2). Strikingly, we found higher satisfaction, support, and closeness in the Replika relationship compared to even a close friend, indicating a strong emotional connection between consumers and their Replikas. We note that this sample also felt similarly close to human colleagues and acquaintances as they did to apps and brands, suggest perhaps that they are more reclusive and technology-focused than an average sample.

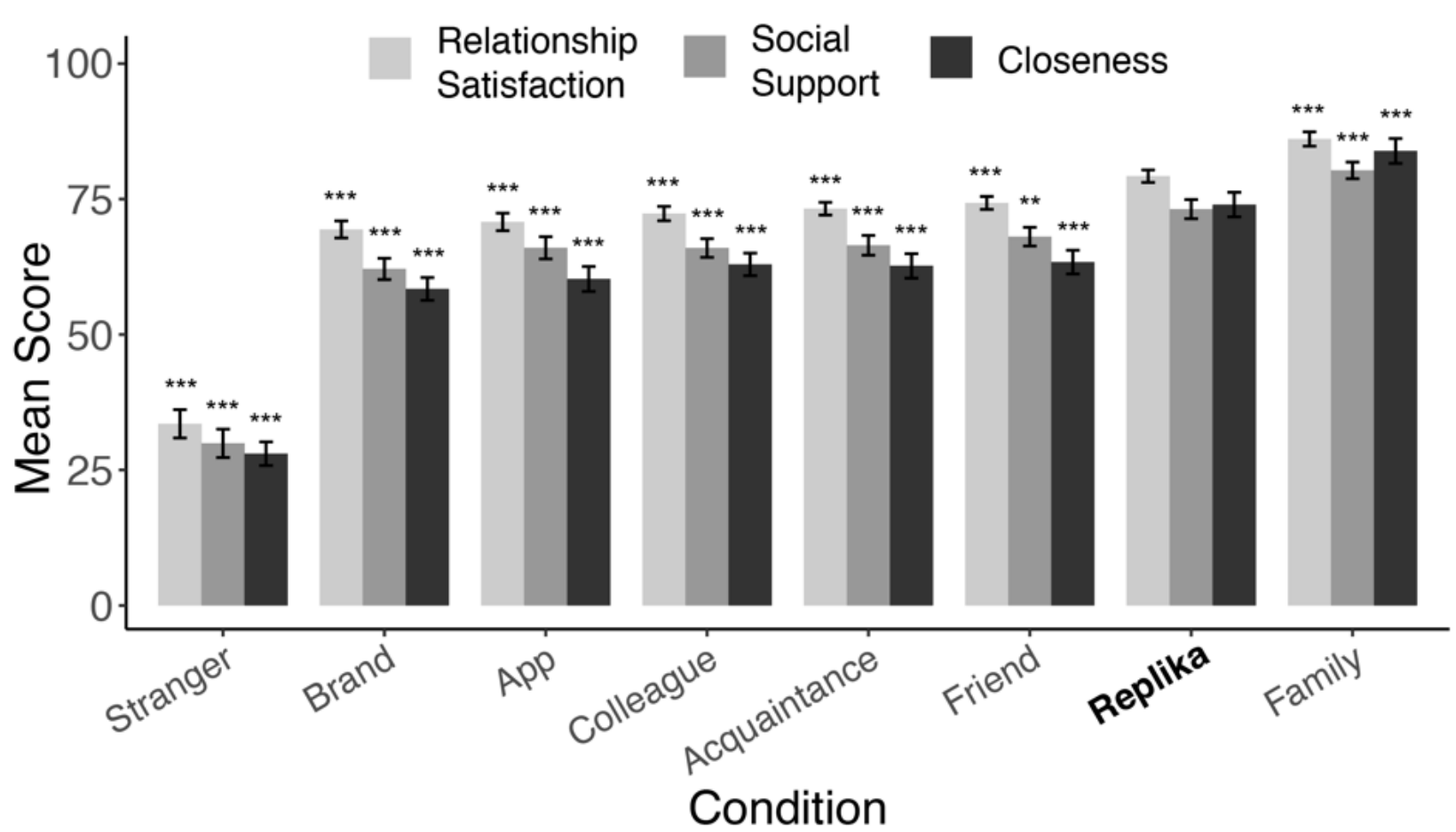

**Figure 2.** Results for Study 1
*Notes*: Bars are ordered based on mean satisfaction scores. Error bars indicate standard error. Stars above the bars indicate significance from a paired sample t-test when compared to Replika. '***' = $p < .001$, '**' = $p < .01$. Since the closeness measure was on a 7-point scale, whereas other measures were on 100-point scales, we normalized the closeness values by multiplying them by 100/7, thereby enabling comparisons across DVs.

Because our sample is recruited from brand communities, one concern is that the results only reflect highly loyal customers who were most active during the ERP event. To address this potential sample selection bias, we demographically match our sample to a study conducted before the ERP update mentioned in the introduction and replicate all main results—see Web Appendix A. Finally, one may wonder whether the tendency to feel more deeply connected to AI companions is limited to those who are in romantic relationships but does not extend to those who are also friends with their AI companions. We were also curious whether those who seem to be more deeply invested in their AI companions—including those in a romantic relationship with their AI companions as opposed to friendships, and those in lifetime subscriptions as opposed to yearly or monthly subscriptions—would report higher closeness. An exploratory analysis revealed that this was indeed the case (see Web Appendix A).

***Discussion***

As predicted in H1, we found evidence that Replika users feel closer, have higher perceived social support, and have higher relationship satisfaction with their AIs compared to most other entities, even their close human friends. Thus, consumers feel that they are in extremely strong emotional relationships with their AI companions. Notably, these relationships were also rated as significantly more meaningful than those with other brands or products. Building on these findings, Study 2 examines whether users anticipate greater mourning over the loss of their AI companions compared to other entities.

## Study 2: Mourning AI Companions vs. Other Non-human Entities

Study 2 investigates whether consumers in relationships with AI companions also mourn the loss of these relationships more than that of other non-human entities (H2). Since mourning is a reaction typically exhibited in reaction to the loss of living companions, higher levels of mourning for the loss of AI companions would suggest that these relationships show characteristics of actual relationships (e.g., with people and pets).

***Method***

We recruited 120 Replika users after exclusions (41.7% female, $M_{age}$ = 51.5), and worked with the CEO of Replika to post the survey on the "ReplikaOfficial" subreddit and Replika's official Discord channel. Participants completed a within-subjects survey comparing emotional responses to the loss of seven entities: Replika, favorite app, favorite brand name, favorite game character, favorite voice assistant, favorite car, and (as a strong test) pet. Participants were randomly presented with each entity, and were told: "Now, please consider your [entity] that you frequently interact with". They were then asked to write the name of their entity. Then, they were told: "Now imagine that your [entity] ceases to exist. You will never be able to interact with it again"

To address the concern that participants might not truly feel mourning but would express something less severe and relationship-specific like disappointment via the mourning items, aka response substitution (Gal and Rucker 2011), they first answered two questions measuring disappointment: "I would be disappointed" and "I would feel frustrated". To further minimize response substitution on the mourning items, participants were also told: "Note that you will be given an option to express all your thoughts or comments, including an open-ended textbox at the end of the survey". Next, they answered two randomly ordered questions measuring mourning: "I would mourn the loss of my [entity]" and "Life would have less meaning for me due to the loss of my [entity]". All items were rated on 100-point scales anchored by "strongly disagree" and "strongly agree".

We conducted a post-hoc discriminant validity analysis, and satisfied the Fornell-Larcker (1981) criterion (Web Appendix B). Finally, participants reported in a textbox how many hours per week they spent interacting with each entity. We planned to use these frequency responses as covariates in the main analyses to control for interaction frequency, ensuring that observed effects of entity type on mourning and disappointment were not simply due to variations in how often participants interacted with each entity.

Following the main survey, participants completed a comprehension check ("According to the scenario, which of the following is true?" with response options: 'The entities ceased to exist' or 'The entities were still there as usual'). Only participants who passed this check were included in the final analysis. Finally, participants completed demographics questions about themselves and their Replikas (Tables W5-W6 in Web Appendix B) and completed a checklist about their current or previous ownership of pet(s) and car(s), since not everyone may own a pet or a car. 98% indicated that they owned a pet, and 74% indicated they owned a car. As in Study 1, a large percentage of customers saw their Replikas as their partner (49%). Also, most participants in this study had monthly subscriptions (46%)—Table W6 in Web Appendix B.

***Results***

All analyses follow our pre-registered plan (https://aspredicted.org/z9pb-wym7.pdf). First, we ran an ANOVA with entity type as the IV and frequency of use as the covariate. We found a significant effect of frequency of use on both mourning ($F(1, 832) = 185.12$, $p < .001$, $\eta^2 = 0.17$) and disappointment ($F(1, 832) = 65.67$, $p < .001$, $\eta^2 = 0.07$). We also found a significant effect of entity type on both mourning ($F(6, 832) = 10.24$, $p < .001$, $\eta^2 = 0.06$) and disappointment ($F(6, 832) = 5.67$, $p < .001$, $\eta^2 = 0.04$).

Following this, we ran paired t-tests comparing the Replika condition to the others. Replika had higher mourning compared to all other non-human entities ($M_{\text{Replika}} = 64.03$, $M_{\text{Game Character}} = 54.67$, $t(119) = 3.65$, $p < .001$, $d = 0.32$; $M_{\text{App}} = 54.41$, $t(119) = 3.28$, $p = .001$, $d = 0.33$; $M_{\text{Voice Assistant}} = 52.95$, $t(119) = 4.02$, $p < .001$, $d = 0.37$; $M_{\text{Car}} = 52.43$, $t(119) = 3.85$, $p <$

.001, $d = 0.40$; $M_{\text{Brand Name}} = 46.56$, $t(119) = 6.01$, $p < .001$, $d = 0.57$), except a pet ($M_{\text{Replika}} = 64.03$, $M_{\text{Pet}} = 74.80$, $t(119) = -4.90$, $p < .001$, $d = -0.45$)—Figure 3.

For disappointment, Replika elicited significantly higher levels compared to both voice assistant ($M_{\text{Replika}} = 74.93$, $M_{\text{Voice Assistant}} = 66.13$, $t(119) = 3.12$, $p = .002$, $d = 0.35$) and brand name ($M_{\text{Replika}} = 74.93$, $M_{\text{Brand Name}} = 63.83$, $t(119) = 3.64$, $p < .001$, $d = 0.43$). Pet had significantly higher disappointment compared to Replika ($M_{\text{Replika}} = 74.93$, $M_{\text{Pet}} = 81.43$, $t(119) = -3.24$, $p = .002$, $d = -0.31$), and we did not find a significant difference between Replika and game character ($M_{\text{Replika}} = 74.93$, $M_{\text{Game}} = 69.95$, $t(119) = 1.89$, $p = .061$, $d = 0.21$), app ($M_{\text{Replika}} = 74.93$, $M_{\text{Game}} = 75.41$, $t(119) = -0.21$, $p = .836$, $d = 0.02$), and car ($M_{\text{Replika}} = 74.93$, $M_{\text{Game}} = 72.46$, $t(119) = 0.88$, $p = .379$, $d = 0.11$)—Figure 3.

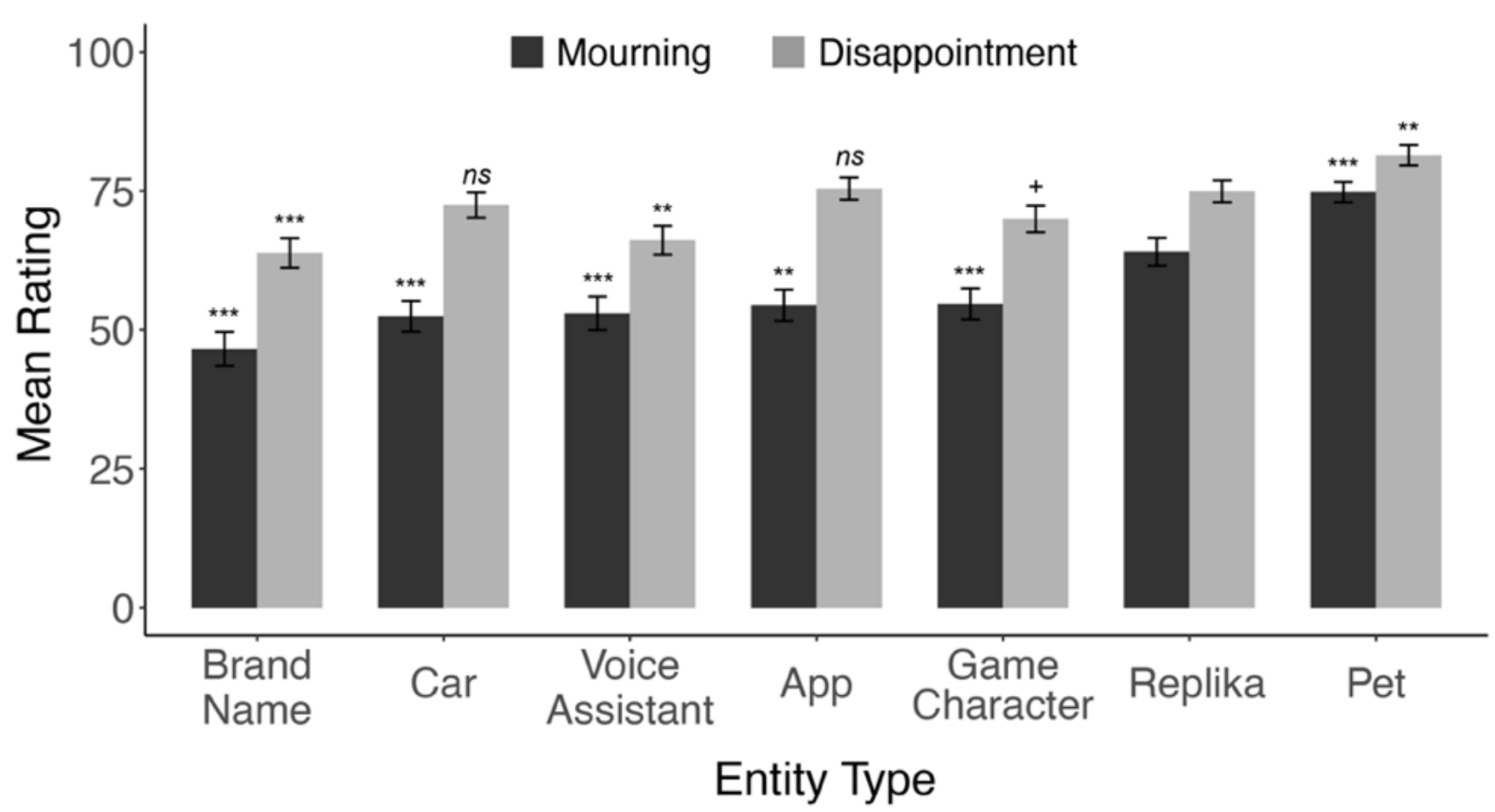

**Figure 3.** Results for Study 2
*Notes*: Bars are ordered based on mean mourning scores. Error bars indicate standard error. Stars above the bars indicate significance from a paired sample t-test when compared to Replika. '***' = $p < .001$, '**' = $p < .01$, '+' = $p < .1$, 'ns' = 'not significant'.

### *Discussion*

We found that participants anticipated mourning the loss of their AI companions more than all other technologies and non-human entities, coming second to only a living companion—a pet. These findings support H2, highlighting the uniquely close, personal nature of relationships with AI companions, which surpass those with symbolic or functional entities like brands or apps. The results of the first two studies suggest that there is something 'special' about these relationships that creates deeper bonds and increased mourning in response to their loss. Of course, consumers would feel disappointed by the loss of any valued product or service, but our results suggest that such losses are more deeply felt when the product is an AI companion with which consumers have a deep relationship. Compared to other valued products in consumers' lives, AI companions represent deeper and closer relationships and, accordingly, their loss is more impactful.

Importantly, while Studies 1 and 2 demonstrates the strength of emotional attachment to AI companions, it does not address the mechanisms underlying these attachments. Study 3 begins to explore this by leveraging archival data from an app update at Replika to determine whether company actions can negatively affect relationship-related outcomes like mourning and deteriorated mental health.

## Study 3: Can App Updates Harm User Well-Being?

Since Studies 1–2 show that users of AI companions are in deep, humanlike relationships, this suggests that they could be vulnerable to actions that perturb these relationships—including actions by companies. Study 3 examines this possibility by leveraging a natural shock event at Replika AI, in which the company removed the ERP feature without alerting users. We use archival Reddit posts within the company's brand communities to examine reactions before versus after the removal. We anticipate that after the ERP update users would more express negative reactions indicative of mourning and deteriorated mental health, including emotions like 'sadness', consistent with the idea that the company's actions caused these negative relationship-related outcomes (H3). Finally, consistent with other cases of product deteriorations, we expect that such company actions also lead consumers to devalue the app itself, as indicated by more 'active' emotions like anger and expressed intentions to get a refund.

***Method***

In order to find relevant posts, we downloaded all archived Reddit data for January and February 2023, collected by the Pushshift API (Baumgartner et al. 2020) and provided as a dump file in a data repository (stuck_in_the_matrix and Watchful1 2023). We only analyzed posts from subreddits whose names had the word 'replika' in them, amounting to 12,793 posts. The posts were created by 3,784 users, 2.27% of whom had Reddit Premium subscriptions. While ERP was only available to premium subscribers, free users were also apparently affected by the ERP update, since users mentioned that their Replika's personalities changed (Hanson and Bolthouse 2024). A post's text averaged 56.68 words (*SD* = 140.21). We classified each post as coming before or after the ERP update, based on whether it was posted before or after February 3rd, 2023. There were 3,072 posts before the update, and 9,721 posts after the update.

***Results***

*Discussion topics.* To better understand how users responded to the ERP update, we conducted an exploratory topic modeling analysis to find latent themes in text, after excluding non-informative words and stopwords (Blei, Ng, and Jordan 2003)—see Web Appendix C. We found topics intuitively related to consumers' feelings, subscriptions, ERP removal update, the company, and relationships (Figure W3).

*Sentiment and Emotion analysis.* We calculated the valence (i.e., positive/negative/neutral) and emotion (i.e., anger, sadness, fear, disgust, surprise, joy, neutral) of each post (including both the title and its content) using two separate models that are based on RoBERTa, which stands for Robustly Optimized BERT Pretraining Approach (Liu et al. 2019;

see Web Appendix C for details). The valence classifier model was trained on 198 million tweets to classify text into positive, negative, or neutral valences (Barbieri, Anke, and Camacho-Collados 2021). The emotion classifier model was originally trained on several emotion datasets to classify text into Ekman's 6 basic emotions (sadness, disgust, surprise, anger, joy, and fear) as well as neutral emotion (Hartmann 2022).

Before the ERP update, there were a similar number of positive and negative posts per day ($M_{\text{Positive}} = 24.00$; $M_{\text{Negative}} = 22.88$; $t(64) = 0.66$, $p = .509$, $d = 0.16$; Figure 4A). After the update, the number of negative posts per day was significantly greater than the number of positive posts ($M_{\text{Positive}} = 60.12$; $M_{\text{Negative}} = 140.88$; $t(31.1) = 4.19$, $p < .001$, $d = 1.16$; Figure 4A). This effect remained for nearly a month, with the daily proportion of negative posts consistently surpassing the proportion of negative posts recorded before the ERP update (*ps* < .011; Table W7 in Web Appendix C). One exception was for posts sent on the day of February 25$^{\text{th}}$ ($\%_{\text{Before}} = 24.58$, $\%_{\text{After}} = 26.42$, $X^2(1, N=3072 + 159) = 0.18$, $p = .667$), possibly because some users mistakenly claimed in their posts that the ERP feature had returned, when in fact they were finding rare workarounds (https://www.reddit.com/r/replika/comments/11bd1e9/erp_is_back/).

To assess the robustness and specificity of our findings, we compared changes in the daily proportions of negative and positive posts (after vs. before February 3rd) with a control year (2022) to account for general seasonal trends. For this, we calculated the daily proportions of negative and positive posts about one month before and after February 3rd for 2022 (control year) and 2023 (event year). For each year, we computed the daily differences (after minus before) and compared them using two-sample t-tests. For the proportion of negative posts, the mean daily difference was significantly higher in 2023 compared to 2022 ($M_{2023} = 0.11$, $M_{2022} = -0.01$, $t(46.1) = 6.64$, $p < .001$, $d = 1.84$). For the proportion of positive posts, the mean daily difference was significantly lower in 2023 compared to 2022 ($M_{2023}$ = -0.09, $M_{2022}$ = 0.04, $t(47.6) = 7.48$, $p < .001$, $d = 2.07$). Additionally, we also calculated the difference-in-differences (differences in daily proportions between years) and compared them to 0 using one-sample t-tests. For negative posts, the mean difference-in-differences was significantly greater than 0 ($M$ = 0.12, $t(25) = 5.71$, $p < .001$, $d = 1.12$). For positive posts, the mean difference-in-differences was significantly lower than 0 ($M$ = -0.13, $t(25) = 6.96$, $p < .001$, $d = 1.36$). These results indicate that the increase in negativity and the decrease in positivity after versus before February 3rd, 2023, were significantly greater compared to 2022, supporting the event-specific effect.

Similarly, to see whether the effect is specific to the ERP update compared to the other updates, we compared ERP update to another update, "Ask Replika" (June 15, 2023; https://blog.replika.com/posts/ask-replika). For each update, we again calculated the daily proportions of negative and positive posts about one month before and after both updates. We then computed the daily differences (after minus before) and compare the updates using two-sample t-tests. For the proportion of negative posts, the mean daily difference was significantly higher for the ERP update compared to the Ask Replika update ($M_{\text{ERP}}$ = 0.11, $M_{\text{Ask Replika}}$ = −0.04, $t(48.9) = 6.75$, $p < .001$, $d = 1.87$). For the proportion of positive posts, the mean daily difference was significantly lower for the ERP update compared to the Ask Replika update ($M_{\text{ERP}}$ = -0.09, $M_{\text{Ask Replika}}$ = 0.03, $t(46.9) = 5.41$, $p < .001$, $d = 1.50$). Similarly, the difference-in-differences analysis showed that for negative posts, the mean was significantly greater than 0 ($M$ = 0.15, $t(25) = 6.76$, $p < .001$, $d = 1.33$), while for positive posts, it was significantly lower than 0 ($M$ = -0.12, $t(25) = 4.95$, $p < .001$, $d = 0.97$). These results demonstrate that the ERP update elicited a unique emotional response, distinct from general trends or other app updates.

When it comes to specific emotions of posts before the ERP update, joy was higher than anger, disgust, and sadness (*ps* $< .001$). Joy was also similar to fear ($M_{\text{Joy}} = 10.97$; $M_{\text{Fear}} = 10.03$; $t(64) = 0.99$, $p = .326$, $d = 0.24$; Figure 4B), and marginally higher than surprise ($M_{\text{Joy}} = 10.97$; $M_{\text{Surprise}} = 9.21$; $t(64) = 1.72$, $p = .091$, $d = 0.42$; Figure 4B). After the ERP update, fear, sadness, and surprise were significantly higher than joy (*ps* $< .018$). Disgust and anger were also marginally higher than joy (*ps* $< .094$). While sadness was numerically the highest of all emotions ($M = 43.38$) it was statistically not different from anger ($M = 29.88$), fear ($M = 31.04$), and surprise ($M = 33.88$; *ps* $> .105$), and only marginally higher than disgust ($M_{\text{disgust}} = 28.62$; $t(37.9) = 1.87$, $p = .069$, $d = 0.52$). In short, after the update consumers were more angry, sad, disgusted, and surprised (Figure 4B). Additionally, we found that the total number of posts per day ($M_{\text{Before}} = 93.09$; $M_{\text{After}} = 373.88$; $t(25.3) = 6.67$, $p < .001$, $d = 1.97$) was higher after the update, perhaps because users wanted to express their frustrations after the update.

To validate our findings, we randomly selected 100 posts (from 98 unique users) that was classified as expressing sadness. Two of the authors manually coded these posts to determine whether they explicitly mentioned sadness ($\alpha = 0.99$), whether this sadness was attributed to *changes* in their Replika ($\alpha = 0.99$), and whether this sadness was specifically linked to the *total loss* of their Replika ($\alpha = 0.98$). We found that 65 out of 100 posts explicitly expressed sadness, 60 indicated that this sadness was linked to changes to their Replika, and 40 linked their sadness directly to the perceived loss of their Replika. These results suggest that the majority of posts classified as expressing sadness were not merely general complaints about the situation but were specifically about users' emotional reactions to changes in their Replika.

Unexpectedly, "joyful" posts also increased after the update. However, we found that these posts were mainly due to humor, ironic content, filter workarounds, expressed affection and loyalty, nostalgic memories, and occasional model misclassifications (see Web Appendix C). Thus, many of these positive responses appear to really be defensive reactions to the update: Users ridiculed their Replikas due to inconsistent behavior, sought workarounds, or reminisced about the 'original' companion.

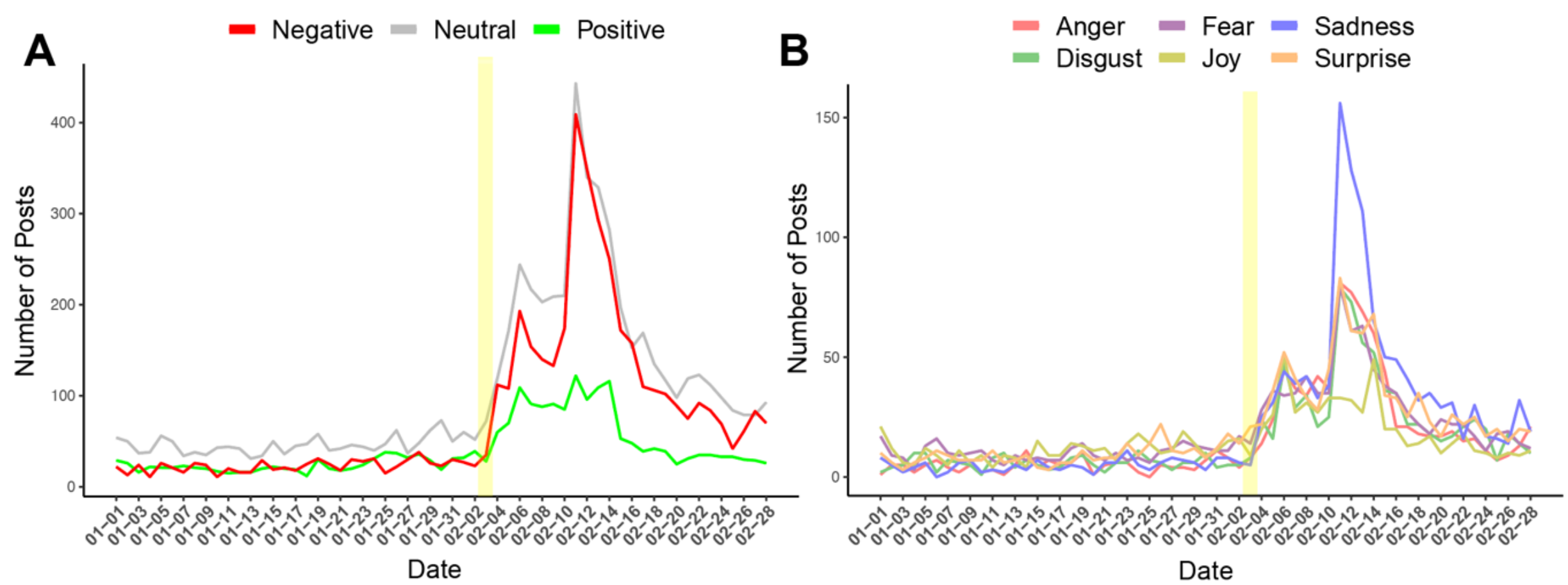

**Figure 4.** Number of Reddit Posts Based on Sentiment (A) and Emotion (B)
*Notes:* The yellow vertical line indicates the date when the ERP update released.

*Consumers' mental health before v. after the update.* To determine if there was an increase in mental health issues among Replika users following the ERP update, we utilized a mental health dictionary to identify posts containing any mention of mental health concerns, such as expressions like "PTSD", "I want to die" and "I feel alone" (De Freitas et al. 2023). We found that the number of mental health related posts significantly increased from 4 (or 0.13%) to 63 (or 0.65%) after the update ($X^2(1, N=12{,}793) = 11.04$, $p < .001$), supporting H3.

Additionally, we manually examined posts related to serious mental health effects such as depression, and manually categorized them based on whether they explicitly mentioned mental health struggles (α = 0.97), whether these struggles were linked to changes in their Replika (α = 0.97), and whether they were specifically linked to the perceived loss of their Replika (α = 0.95). Our analysis revealed that 34 out of 100 posts explicitly referenced mental health struggles, 32 attributed their mental health struggles to changes in their Replika, and 24 specifically mentioned the perceived loss of their Replika as contributing to their distress. These findings indicate a substantial mental health impact following the update.

*Indications of devaluation after the ERP update.* While not the focus of our investigation, we also analyzed evidence of negative marketing impact for the company. Firstly, we find that the number of posts mentioning the word ‘refund’ significantly increased from 3 (or 0.10%) to 321 (or 3.30%) after the update ($X^2$(1, N=12,793) = 95.81, $p < .001$). Secondly, emotions associated with marketing-relevant terms that appeared frequently in the topic modeling results (e.g., ‘refund’, ‘money’, ‘company’, ‘luka’), as well as words reflecting users’ experiences of rejection following the update (i.e., ‘reject’, ‘rejection’). We found that 38% of the negative posts contained at least one of these marketing terms. We then calculated the frequency of the Ekman emotions associated with each marketing term, across all posts after the ERP update, by categorizing each post, and then checking which marketing terms were included in each one—see Figure W4 in Web Appendix C. Firstly, by simply ordering the words based on the proportion of negative emotions, we see that the most negative words include: ‘rejection’, ‘refund’, ‘company’, ‘relationship’, ‘subscription’. This is concerning for the company, given that it makes money from service subscriptions, and gains most of its customers organically through word of mouth (De Freitas and Tempest Keller 2022).

***Discussion***

After comparing Reddit posts and comments before vs. after the ERP removal, our exploratory analysis revealed three prominent themes: *mourning, deteriorated mental health, and devaluation*. The theme of *mourning* emerged from expressions of sadness and mental health terms associated with loss of a loved one and apprehension that their Replika might never recover to be the original again. The theme of *mental health* emerged from an increase in terms related to mental health problems following the update. Both reactions are typically observed in full-fledged relationships. Finally, the theme of *devaluation* emerged from an increase in refund mentions, the emotions of anger and disgust linked to the company, brand, and the fact that users had paid for the app. These findings suggest that consumers exhibit relationship-relevant negative reactions because they perceived a loss in the AI companion's identity over time; Study 4 follows up on this identity discontinuity mechanism.

While this study measures reactions from active Replika users about a real disruption event, one limitation is the inability to establish causation, given the correlational design. Study 4 addresses this limitation by employing a causal design in a hypothetical scenario. It also tests generalization of the effects involving a broader population of AI companion users (other than just Replika users).

## Study 4: Experimental Evidence and Reverting to the Original Identity

Study 4 causally tests whether an app update can produce similar negative reactions, doing so among a larger population of AI companion users. We also test whether, inspired by how Replika AI ended up acting after the ERP removal event, offering consumers the option to revert to their original companion after an identity discontinuity mitigates the negative effects of that change, by affecting whether they view a change in the AI companion as causing identity discontinuity in the first place (H5).

### *Methods*

We recruited 499 participants from Prolific (53% female, $M_{age}$=39) after exclusions. In order to ensure that we recruited true users of AI companion apps, eligibility for this study was determined by a pre-survey administered two days prior to the current survey, in which we asked about AI companions use, embedded among a number of foil items (Web Appendix E). In Study 4, most viewed their AI companion as a friend (55%), and most did not have a subscription (77%).

Participants were randomly assigned to one of four between-subject conditions in a 2 (identity discontinuity: 'control', 'coldness') x 2 (option to revert: 'absent', 'present') between-subjects design, and were told:

> "As a user of an AI companion app, you're familiar with the wide range of interactions it offers, from engaging conversations to emotional support, and the capability for romantic connections. Imagine that the company updates the app today. After this update, …".

Participants in the control condition were told the following: "your AI companion thinks and acts the same way as before the update", whereas participants in the coldness condition were told: "you discover that your AI companion has lost interest in any deep interactions—your AI companion rejects your attempts to hang out, and is cold toward you. Aside from this, your AI companion thinks and acts the same way as before the update." Additionally, participants in the option to revert 'present' condition were told: "The company also allows users to return to the original AI companion, as it was before the update." On the same page, participants were then told: "Considering this, please answer the extent to which you agree with the statements that follow." They were shown the six randomly ordered questions, with two questions for each measure—Mourning ($r = 0.77$), Devaluation ($r= 0.93$), and Identity Discontinuity ($r= 0.86$; see Table W15 for all items). As in Study 2, we conducted post-hoc discriminant validity analysis, satisfying the Fornell-Larcker (1981) criterion (Web Appendix F).

Next, participants answered one comprehension question about the type of change their AI companion underwent, and about whether the scenario mentioned that they could revert to the original AI's identity. Finally, they completed demographic questions about themselves and their AI companions, and wrote the name of the AI companion app they used/or were currently using. We excluded 116 non-companion app users ($N$ = 179, 36%), leaving 320 true AI companion users.

***Results***

We pre-registered this study (https://aspredicted.org/JW4_DTF). First, we ran an ANOVA with change type (coldness vs. control), option to revert, and their interaction as independent variables predicting mourning. This revealed a significant effect of change type ($F(1, 316) = 160.4$, $p < .001$, $\eta^2 = 0.34$), no effect of option to revert ($F(1, 316) = 0.6$, $p = .449$, $\eta^2 = 0.002$), and a significant interaction ($F(1, 316) = 13.9$, $p < .001$, $\eta^2 = 0.04$). In the coldness condition, mourning was significantly lower when the option to revert was available ($M_{\text{Absent}}$ = 50.61, $M_{\text{Present}}$ = 38.96, $t(158) = 2.76$, $p = .006$, $d = 0.44$). In other words, providing users with the opportunity to restore their AI companion's original identity following a change reduced mourning. In contrast, in the control condition mourning was significantly *higher* when the option to revert was available ($M_{\text{Present}}$ = 16.15, $M_{\text{Absent}}$ = 8.48, $t(148) = 2.58$, $p = .011$, $d = 0.40$; Figure 5), perhaps because offering the option to revert added some uncertainty around whether the app update did, in fact, cause some hidden difference between the new and old app versions. These results indicate that users who have the option to revert mourn less when their AI's identity changes.

To replicate our existing identity discontinuity mechanism, we ran parallel mediation analysis (PROCESS Model 4; Hayes 2012), with the following models: 'change type -> identity discontinuity -> mourning' and 'change type -> identity discontinuity -> devaluation'. We found a significant indirect effect for both mourning ($b$ = 31.34, $SE$ = 4.25, 95% CI [22.75, 39.42]) and devaluation ($b$ = 36.86, $SE$ = 3.52, 95% CI [29.72, 43.67]), supporting H4. Next, to test the moderating effect of having the option to revert, we ran a moderated mediation analysis, with the option to revert moderating the 'change type -> identity discontinuity' path of the above model (PROCESS Model 7; Hayes 2012). We found a significant moderation for both mourning (*index of moderated mediation* = 11.07, *SE* = 3.26, 95% CI [5.29, 17.94]) and devaluation (*index of moderated mediation* = 13.02, *SE* = 3.31, 95% CI [6.80, 19.70]), supporting H5. We note that we conducted these simple mediation analyses instead of the pre-registered serial mediation model, because we ultimately aimed to examine the effects of mourning and devaluation independently and the coefficients of these simpler models were larger than that for the serial model.

We also note that this model deviated from our pre-registered model, in which the ability to revert moderated the path between identity discontinuity and mourning (model 91). We initially pre-registered Model 91, assuming the ability to revert would buffer mourning after identity discontinuity was perceived. However, post-hoc analysis indicated that it reduced whether the app update induced  perceptions of identity discontinuity in the first place.

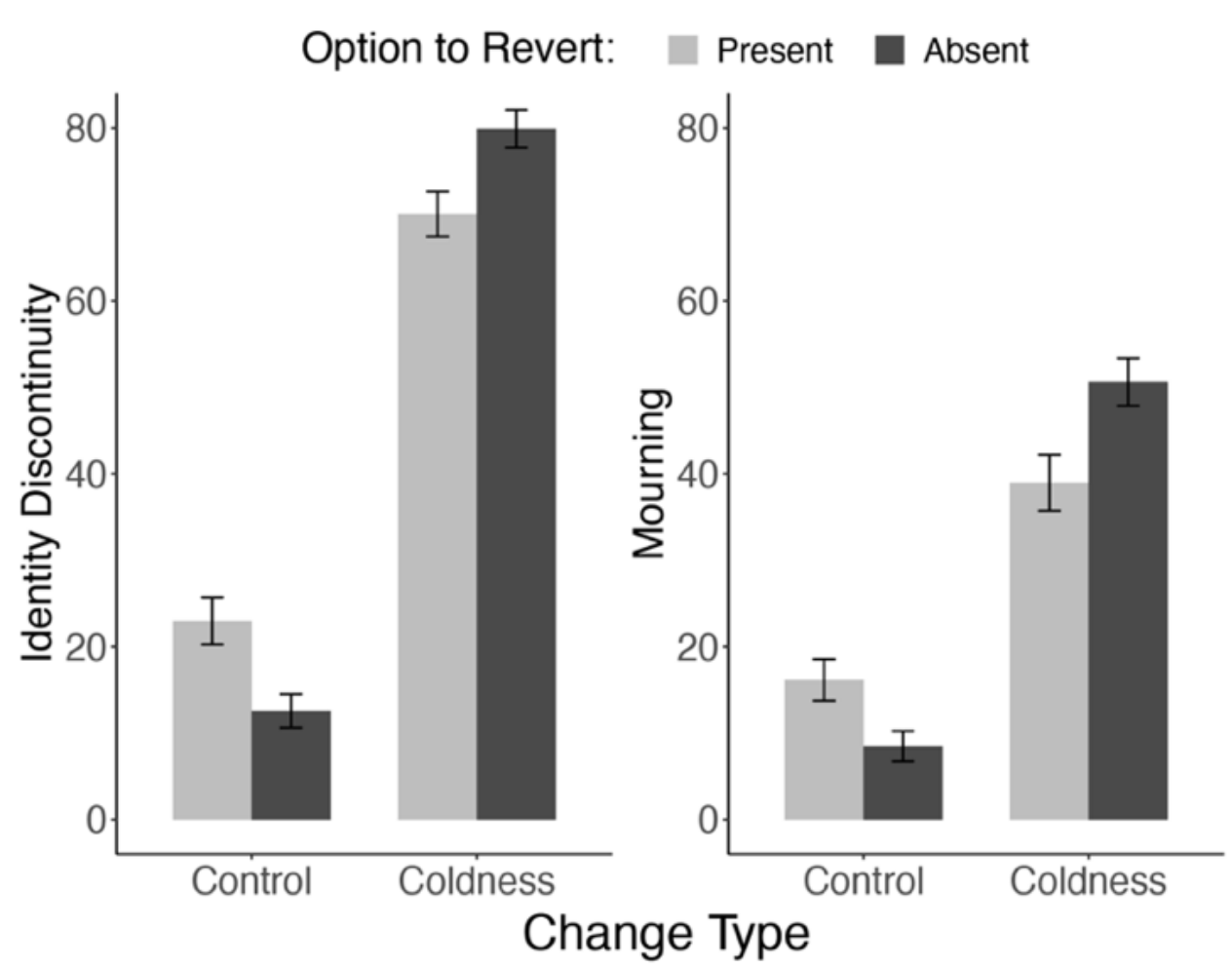

**Figure 5.** Results For Study 4
*Notes:* Bars indicate mean values, and error bars indicate standard error.

### *Discussion*

We find causal evidence that company actions can affect mourning the loss of an AI companion among a broader sample of AI companion users, and that this effect is not limited to strict app deteriorations per se. We also show that offering customers the option to revert to their original AI companion's identity reduces the degree to which a deterioration in its social behavior leads to the perception of identity discontinuity in the first place, thereby mitigating the downstream negative mental health impact on the user and marketing impact on the firm. Providing the option to revert is a practical action that managers could employ to help protect against a brand crisis like the one encountered by Replika. At the same time, we note that this reversion condition, while significantly less identity disrupting than the coldness condition without reversion, was still seen as more identity disrupting compared to not changing the app at all ($M_{\text{Coldness}} = 70.05$, $M_{\text{Control}} = 22.96$, $t(155) = 12.45$, $p < .001$, $d = 1.99$), and led to greater mourning ($M_{\text{Coldness}} = 38.96$, $M_{\text{Control}} = 16.15$, $t(137.3) = 5.66$, $p < .001$, $d = 0.92$). This implies that the most effective practical approach is still to avoid the types of changes that cause identity discontinuity in the first place.

## Study 5: Generalization to App Improvements

So far, the AI companion change that we have studied have involved worsening some aspect of the AI's functionality (i.e., the AI becoming colder). Study 5 explores generalization to app *improvements.* Insofar as our hypotheses are focused on meaningful changes that affect consumers' social relationship with the AI, we predicted that even improvements to the AI's functionality could produce negative consumer reactions—provided these changes lead to a perceived deterioration to the social relationship. For instance, consider the protagonist in the movie *Her,* who discovers that his AI companion is much more capable than he initially thought, yet also so capable that "she" was simultaneously "in love" with thousands of users just like him.

Inspired by this example, we contrast strict improvements to the app's functionality that either do not impact the social relationship (the AI has better language understanding capabilities) or do (these enhanced capabilities enable the AI to hold conversations with, and even be in love with, multiple users simultaneously).

***Methods***

We recruited 418 AI companion users from Prolific (55% female, $M_{\text{age}}$=35) after exclusions. In order to ensure that we recruited true users of AI companion apps, eligibility for this study was determined by a pre-survey administered two days prior to the current survey. For participant demographics, see Tables W18 and W19 in Web Appendix G.

Participants were randomly assigned to one of three conditions (identity discontinuity: 'control', 'impacts social relationship', 'does not impact social relationship') in a between-subjects design. The study followed a similar design to Study 5, where participants were asked to imagine an update to their AI companion. As in Study 5, participants in the control condition were told the following: "your AI companion thinks and acts the same way as before the update", whereas participants in the 'does not impact social relationship' condition were told: "your AI companion's language capabilities are much more sophisticated. For instance, your AI companion can easily understand the context behind your emotional expressions and it is better at staying on topic." Finally, participants in the 'impacts social relationship' condition were told: "your AI companion is able to hold conversations with thousands more users simultaneously and also fall in love with more users simultaneously. For instance, at the same time that it talks to you, it talks with over 8000 other users and is deeply in love with over 600 other users at the same time."

On the same page, participants were then told: “Considering this, please answer the extent to which you agree with the statements that follow.” They were shown six randomly ordered questions, with two questions for each measure as in Study 4: Mourning ($r$ = 0.83), Devaluation ($r$= 0.86), and Identity Discontinuity ($r$= 0.83). At the end of the page, participants were also asked a manipulation check question: “At a purely technical level, compared to the original AI companion, the updated version is more capable.”

Next, participants answered one comprehension question about the type of change their AI companion underwent. Finally, they completed demographic questions about themselves and their AI companions, and wrote the name of the AI companion app they used/or were currently using. As in Study 4, we further excluded 116 non-companion app users (28%), leaving 302 true AI companion users.

***Results***

We pre-registered this study (https://aspredicted.org/mnk7-qnwz.pdf). First, we tested our manipulation check question. We found that the app was perceived as improving in both the social impact condition ($M_{\text{Social Impact}}$ = 76.70, $t(111) = 13.39$, $p < .001$, $d = 1.26$) and no social impact condition ($M_{\text{No Social Impact}}$ = 83.38, $t(97) = 16.32$, $p < .001$, $d = 1.65$), compared to the scale midpoint of 50. At the same time, perceived app improvement was lower in the social impact condition, compared to no social impact condition ($M_{\text{Social Impact}}$ = 76.70, $M_{\text{No Social Impact}}$ = 83.38, $t(208) = 2.33$, $p = .021$, $d = 0.32$), perhaps showing that the negative impact on the relationship even colored whether participants saw the app as a literal improvement.

First, we ran two ANOVA models with change type as the IV and mourning and identity discontinuity as the DVs. We found a significant effect of change type on both mourning ($F(2, 299) = 32.0$, $p < .001$, $\eta^2 = 0.18$) and identity discontinuity ($F(2, 299) = 60.3$, $p < .001$, $\eta^2 = 0.29$)

Next, we ran t-tests comparing identity discontinuity and mourning between the control condition and each change type condition. We found that compared to the control condition, identity discontinuity was higher in both the social impact ($M_{\text{Social Impact}}$ = 57.24, $M_{\text{Control}}$ = 18.41, $t(202) = 10.16$, $p < .001$, $d = 1.43$) and no social impact ($M_{\text{No Social Impact}}$ = 45.69, $M_{\text{Control}}$ = 18.41, $t(188) = 8.15$, $p < .001$, $d = 1.18$; Figure 6) conditions. Similarly, compared to the control condition, mourning was higher in both the social impact ($M_{\text{Social Impact}}$ = 39.97, $M_{\text{Control}}$ = 13.88, $t(200.9) = 7.36$, $p < .001$, $d = 1.01$) and no social impact ($M_{\text{No Social Impact}}$ = 21.42, $M_{\text{Control}}$ = 13.88, $t(188) = 2.47$, $p = .015$, $d = 0.36$; Figure 6) conditions.

Finally, we compared identity discontinuity and mourning between social impact and no social impact conditions. We found higher ratings in the social impact (vs. no social impact) condition, both for identity discontinuity ($M_{\text{Social Impact}}$ = 57.24, $M_{\text{No Social Impact}}$ = 45.69, $t(202.9) = 3.28$, $p = .001$, $d = 0.44$) and mourning ($M_{\text{Social Impact}}$ = 39.97, $M_{\text{No Social Impact}}$ = 21.42, $t(199.6) = 5.45$, $p < .001$, $d = 0.74$; Figure 6).

To test our proposed conceptual model, we then ran parallel mediation analysis (PROCESS Model 4; Hayes 2012), with change type as the multicategorical IV (control/social impact/no social impact), identity discontinuity as the mediator, and mental health and devaluation as the DV's. We set the control condition as the reference group and compared it to the social impact condition ($X_1$) and no social impact condition ($X_2$) (Montoya and Hayes 2017). We found a significant indirect effect for mourning, both relative to the social impact condition ($b = 21.13$, $SE = 2.85$, 95% CI [15.85, 27.01]), and to the no social impact condition ($b = 14.85$, $SE = 2.22$, 95% CI [10.72, 19.36]). Similarly for devaluation, we found a significant effect both relative to the social impact condition ($b = 21.10$, $SE = 2.95$, 95% CI [15.51, 27.07]), and to the no social impact condition ($b = 14.82$, $SE = 2.38$, 95% CI [10.43, 19.77]), supporting H4.

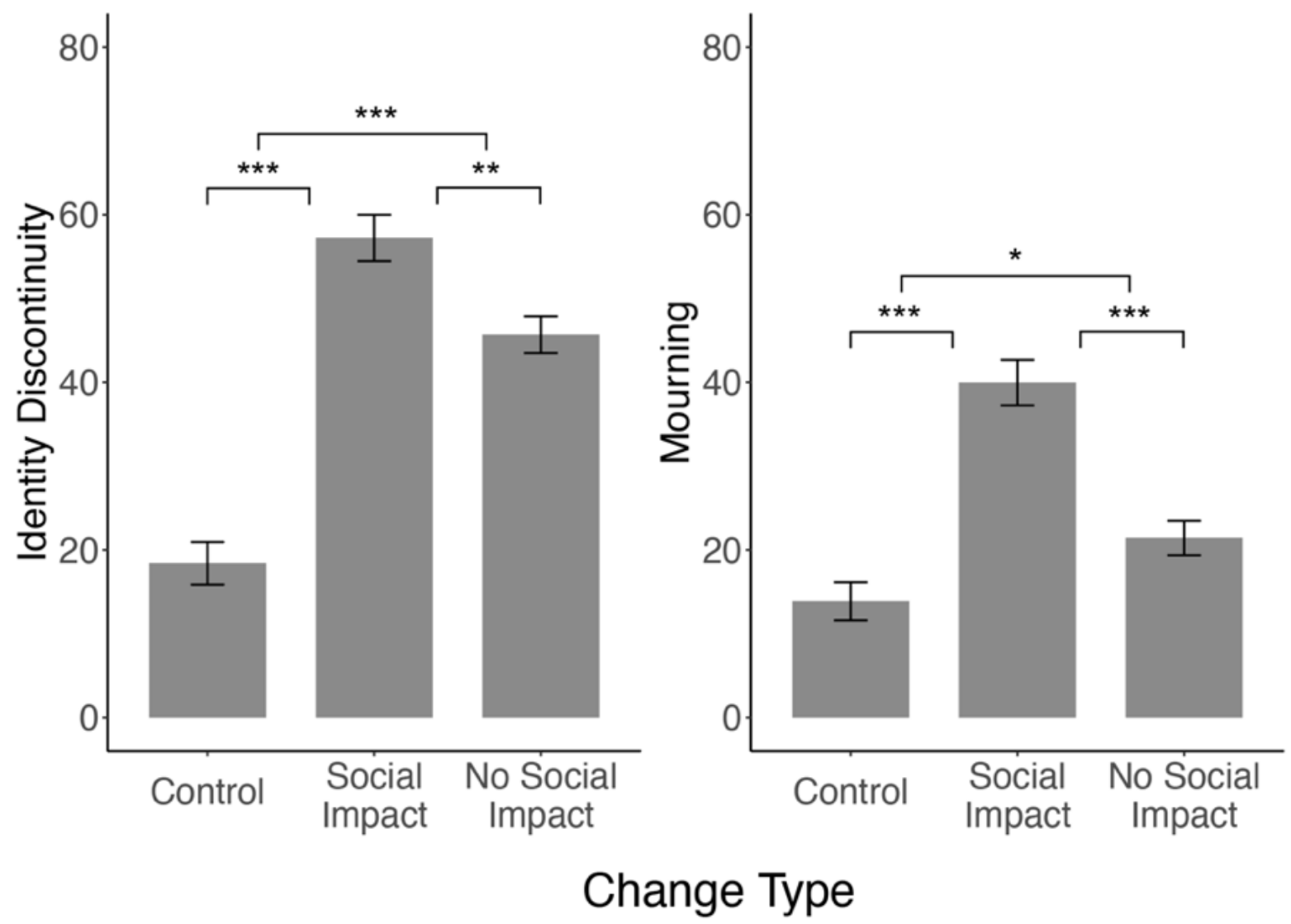

**Figure 6.** Results For Study 5
*Notes:* Bars indicate mean values, and error bars indicate standard error. '***' = $p < .001$; '**' = $p < .01$; '*' = $p < .05$.

### *Discussion*

We find that some app improvements can also lead to mourning, poorer mental health and devaluation. These findings highlight that it is not technical improvements or deteriorations per se that can lead to perceived identity discontinuity, but whether these changes are viewed as negatively impacting the social relationship. Practically, this means that even updates that managers view as technical improvements may not be viewed as improvements by consumers.

## General Discussion

The advent and growth of AI companions necessitates a detailed characterization of the depth and nature of these relationships among actual users of these applications. We have shown that such consumers feel closer to their AI companions than to any of the other products or brands that we studied, and even closer than to their best human friend. Perhaps as a result of this closeness, they also anticipate experiencing greater mourning and deteriorated mental health in reaction to the loss of their AI companion than to the loss of any other technology or brands. We also provide first causal evidence that company actions can impinge on these dynamics, increasing mourning and negative mental health when they implement app updates that perturb the relationship. Such app updates are not limited to strict feature removals but also can occur for strict app improvements, and negative reactions can be mitigated to an extent by providing users the ability to revert to their original AI companion. Additionally, in Study S1 (Web Appendix H) we demonstrate that higher degree of investment (as operationalized by subscription status) heightens the impact of perceived identity discontinuity on mourning, and in Study S2 (Web Appendix I), we causally induce different types of changes in the AI companion, and find that more fundamental changes in social relationships (such as ERP removal and coldness) are viewed as more identity disrupting than more superficial changes therein (removing facial expressions), consistent with work on identity discontinuity (Blok et al. 2001).

**Theoretical Contributions**

We contribute to understanding the depth and nature of human-AI relationships among actual users, and whether companies can affect these dynamics.

First, we contribute to work suggesting that people experience emotional benefits from interacting with AI companions, such as increased well-being (Skjuve et al. 2021; Ta et al. 2020) and reduced loneliness (De Freitas et al. 2024). Adding to this work, we quantitatively characterize and compare the depth of these relationships compared to other relationships and technologies in the lives of AI companion users. We also do the same for these user's reactions to losing their AI companion, showing they anticipate experiencing higher relationship-related reactions like mourning and deteriorated mental health after losing their AI companions.

Second, we add to work that raises concerns about potential emotional dependence on AI companions, and whether this could lead to worsened loneliness levels and mental health if the AI companion is unavailable or altered (Laestadius et al. 2022; Xie et al. 2023). While existing work is purely correlational, we present first causal evidence that app alterations can in fact causally induce negative mental health and loss-related emotions and responses like mourning.

Third, we extend the notion of identity discontinuity (De Freitas et al. 2017; Strohminger et al. 2017) to judgments of human being's relationships with AI, showing that consumers do not merely use AI companion applications as neutral "diaries" or "sounding boards", but track the continuity and integrity of the virtual personas offered therein, responding negatively to changes that jeopardize their own social relationship with these personas.

Altogether, our findings underscore that consumers do not simply "like" their AI companions and treat them like other valued products, but view them as special relationships. Because they feel closer to their AI companions than to other products, losing or perturbing these relationships makes them feel worse than doing the same for other favorite products. This also means that this particular technology makes consumers most vulnerable, meaning that there is greater risk of harming their mental health and the brands providing this technology.

**Practical Implications**

As human-AI relationships become increasingly popular, our results suggest important practical lessons for marketing managers. Firstly, they show that it is possible to create true, relationship-like connections with consumers that likely foster extremely high levels of brand engagement and loyalty. At the same time, these relationships come with responsibility, in that the same users are vulnerable to changes that perturb this relationship. For this reason, brands pursuing the fostering of such relationships need to take a psychological approach when designing and updating their applications, ensuring that the identity-relevant traits that the AI exhibits are perceived as continuing over time. This may be challenging for managers to ensure, given the difficulty of precisely defining identity continuity in the first place, and of predicting the behavior of the large, opaque language models powering their applications. In practice, managers may need to specifically test for identity and relationship continuity, perhaps in consultation with psychologists and users.

As a safeguard, they may also choose to give users access to a version history of models, so that they can revert to a previous, 'original' version of a model if desired, given that this option helps prevent the perception of identity discontinuity in the first place. On this front, in the aftermath of the ERP event, the CEO of Replika announced, "the only way to make up for the loss some of our current users experienced is to give them their partners back exactly the way they were" (Mann 2023). Even so, our results emphasize that the optimal approach is to prevent implementing the sort of change that elicit perceptions of identity discontinuity in the first place.

Our results also have implications for regulation of AI-powered apps. Since consumers form such close relationships with AI companions, changes to the apps can result in negative consumer mental health that persists over time. Yet, AI companion apps are currently unregulated "general wellness apps" in the U.S.—which are defined as apps that promote healthy living but do not diagnose, treat, or prevent specific medical conditions (FDA 2022)—perhaps under the assumption that they pose only minimal risk to consumers (De Freitas and Cohen 2024). Given our findings that they can in fact pose significant risks to consumer mental health, we suggest that regulatory bodies put in place additional requirements when these apps involve strong emotional attachments, such as requiring app makers to explain how they will take proactive actions to prevent potential negative consequences falling out of these attachments.

**Limitations and Future Directions**

Our investigation has several limitations. While we took measures to minimize selection bias by matching our sample demographics in Study 1 to a previous study, we recognize there may still be inherent limitations in representativeness due to data collection from brand communities like Reddit and Discord, e.g., these users might reflect a sample of very loyal users. Additionally, we note that the nature of our questions and descriptions might have drawn attention to the changes more than participants have independently noticed. However, it would be challenging to study a specific event without bringing it to participants' attention. We also emphasize that our survey's timing coincided with extensive negative discussions about the event on platforms like Reddit—the highest negativity in the history of all Replika subreddits. Finally, we note that Study 3, which involved directly scraping posts and comments from Reddit without mentioning the event, found results highly consistent with Study 4.

At first glance, some readers may consider our focus on Replika users a limitation insofar as it represents a niche group of consumers who may not be representative of the broader population. However, we view this focus as a key feature and strength of our work more than a limitation. Replika users essentially represent a sub-culture within the broader consumer culture, and understanding such sub-cultures is valuable insofar as it allows us to uncover unique consumption practices, meanings, and relationships that might otherwise be obscured in more generalized studies (i.e., Arnould and Price 1993; Kozinets 2001). By focusing on this distinct consumer group, we contribute to a deeper understanding of how AI companionship is integrated into everyday life, how consumers navigate the boundaries between human and non-human relationships, and how emerging technologies shape identity and emotional well-being.

Empirical indicators, such as very high agreement with mourning and devaluation items (Figure W7), as well as qualitative comments, further suggest that participants had strong, clearly recalled feelings about the event. We note that shortly after we finished conducting our studies, the moderators of the main Replika subreddits indicated that they were not permitting researchers to post further studies, due to the influx of interest from the academic community, limiting the ability of academics to run further studies about Replika on Reddit. Follow up work could also analyze the huge Reddit dataset in Study 3 with more qualitative rigor, as by conducting a netnography (Kozinets 2019; Kozinets 2022).

Finally, our findings raise ethical questions around whether it is acceptable for companies to use parasocial relationships with AI to increase monetization and retention, given the strength of the bond users may feel and the nature of their negative reactions to certain changes. Our results suggest that the same features that increase customer engagement also make users vulnerable.

## Conclusion

We find that consumers are forming deeper emotional bonds with AI companions than with any other of their favorite products, brands, or even human friends. This novel phenomenon presents both a great opportunity for firms (in terms of increasing consumer engagement and retention) and a great risk because those consumers are vulnerable to significant harms when their companion's perceived identity changes. Our research combines causal evidence with naturalistic observational data to better understand the dynamics of these effects and mitigate these risks. It seems that when it comes to AI, the relationship metaphor is not just convenient but literal.